\documentclass[a4paper,usenatbib]{mnras}
\usepackage{newtxtext,newtxmath}

\usepackage[T1]{fontenc}
\usepackage{ae,aecompl}

\usepackage{graphicx}	
\usepackage{amsmath}	
\usepackage{amssymb}	
\usepackage{caption}	
\usepackage{multirow}

\title[Pseudo-synthetic calibration with Pyramid WFS]{A new calibration strategy for adaptive telescopes with pyramid WFS}

\author[C. T. Heritier et al.]{
C. T. Heritier,$^{1,2,3,4}$\thanks{E-mail: cedric.heritier@lam.fr}
S. Esposito,$^{3}$
T. Fusco,$^{1,2}$
B. Neichel,$^{1}$
S. Oberti,$^{4}$\\
\newauthor
R. Briguglio,$^{3}$
G. Agapito,$^{3}$
A. Puglisi,$^{3}$
E.Pinna$^{3}$
and P.-Y. Madec $^{4}$ 
\\
\\
$^{1}$  Aix Marseille Univ, CNRS, CNES, LAM, Marseille, France\\
$^{2}$ ONERA, DOTA, Unit\'{e} HRA, 29 avenue de la division Leclerc, 92322 Chatillon, France\\
$^{3}$INAF - Osservatorio Astrofisico di Arcetri Largo E. Fermi 5, 50125 Firenze (Italy)\\
$^{4}$European Southern Observatory, Karl-Schwarzschild-str-2, 85748 Garching, Germany\\
}
\date{Accepted 2018 September 06. Received 2018 September 06; in original form 2018 August 27}

\pubyear{2018}

\begin{document}
\label{firstpage}
\pagerange{\pageref{firstpage}--\pageref{lastpage}}
\maketitle

\begin{abstract}
Several telescopes include large Deformable Mirrors (DM) located directly inside the telescope. These adaptive telescopes trigger new constraints for the calibration of the Adaptive Optics (AO) systems as they usually offer no access to an artificial calibration source for the interaction matrix measurement. Moreover, the optical propagation between the DM and the Wave-Front Sensor (WFS) may evolve during the operation, resulting in mis-registrations that highly affect the AO performance and thus the scientific observation. They have to be measured and compensated, for instance by updating the calibration. A new strategy consists of estimating the mis-registrations and injecting them into synthetic models to generate noise-free interaction matrices. This pseudo-synthetic approach is the baseline for the Adaptive Optics Facility working with a Shack-Hartmann WFS and seems particularly suited for the future Extremely Large Telescope as the calibration will have to be regularly updated, for a large numbers of actuators. In this paper, the feasibility of a pseudo synthetic calibration with Pyramid WFS at the Large Binocular Telescope (LBT) is investigated. A synthetic model of the LBT AO systems is developed, and the procedure to adjust the mis-registrations parameters is introduced, extracting them from an experimental interaction matrix. We successfully tested an interaction matrix generated from the model on the real system in high-order AO mode. We recorded a slightly better performance with respect to the experimental one. This work demonstrates that a high accuracy calibration can be obtained using the pseudo synthetic approach with pyramid WFS.
\end{abstract}

\begin{keywords}
instrumentation: adaptive optics - telescopes 
\end{keywords}



\section{Introduction}
Adaptive Optics (AO) is now commonly spread on large aperture optical telescope facilities to compensate in real time the variations of optical index in the atmosphere and retrieve the full angular resolution of the telescope (\cite{babcock1953possibility}).
The principle of a classical AO system is the following: a Wave-Front Sensor (WFS) measures a signal relative to the phase and sends it to a Real Time Computer (RTC) that computes the corresponding commands to apply on a Deformable Mirror (DM). This system is usually operated in a feedback-loop at a higher frequency than the temporal evolution of the turbulence (typically, a few hundred Hz). 
However, to provide a good correction (e.g. to be able to apply the correct shape on the DM at the right time), the AO loop has to be properly calibrated before the operations. This is achieved by measuring the interaction matrix of the system which consists of recording the WFS signals corresponding to a specific command of the DM actuators. 

Recent developments in telescope designs and DM technology have led to consider DM located directly into the telescope ( turning them from active to adaptive telescopes) to reduce the number of optics and increase the numbers of photons available for the instruments. This concept was first validated on the MMT (\cite{wildi2003firstLightMMTsecondaryMirror}) and is now used on two of the largest ground based optical telescopes, the Large Binocular Telescope (LBT), with its two Adaptive Secondary Mirrors (ASM) (\cite{riccardi2003adaptive}, \cite{esposito2010laboratory}) of 672 actuators, and the Unit Telescope 4 of the Very Large Telescope, recently upgraded to become the Adaptive Optics Facility (AOF) (\cite{arsenault2006AOF}, \cite{stroebele2006eso}) with its Deformable Secondary Mirror (DSM) of 1170 actuators.
The next generation telescopes, the Extremely Large Telescope (ELT) (\cite{gilmozzi2007european}) and Giant Magellan Telescope (GMT)  (\cite{johns2006giant}), will also include large adaptive mirrors in their design with respectively 4356 and 4702 actuators (\cite{vernet2012specifications}, \cite{hinz2010gmt}). The Thirty Meter Telescope (TMT) is also considering this design and will thoroughly study this option in the coming months (\cite{boyer2018adaptive}). These numbers put light on the very first challenge for the AO calibration of this new generation of adaptive telescopes: how much of precious telescope time will be required to provide a satisfying calibration of the AO systems?

This problematic was already addressed in \cite{oberti2006large} in which a summary of the main technical challenge is provided: large number of actuators and regular update of the calibration with often no external source illuminating the DM. Indeed, such telescopes usually don't provide any access to an intermediate focal plane ahead of the DM, and thus to any artificial calibration source. This specificity requires to completely rethink the way to acquire the interaction matrix of the system. However, even in presence of such an artificial calibration source, the large distance between the DM and the WFS may frequently affect the registration between both systems (due to gravity or flexure). These so-called mis-registrations\footnote{relative shifts, rotation, magnification or higher order pupil distortion between the DM actuators and WFS subapertures}, affect the performance of the AO correction, especially for high order systems which are extremely sensitive to these mis-alignments and require then a high accuracy calibration. Such calibration errors impact the performance of the scientific instruments and can lead to instability of the AO loop, perturbing the telescope operation. 

The ELT will take these constraints to another level. The first challenge in terms of telescope operation will be to calibrate a large number of actuators with no access to a calibration source. Moreover, the calibration will require regular updates during the operations due to the unprecedented distance between the DM and the WFS with moving optics between them. 

There is then a crucial need to optimise and develop new calibration strategies to overcome these constraints and minimize the telescope time required. Some methods have already been identified to achieve the measurements. The first approach consists of measuring the interaction matrix on-sky (\cite{wildi2004determining}, \cite{esposito2006onSkyIM}, \cite{pieralli2008sinusoidal}, \cite{pinna2012first}, \cite{kellerer2012deformable}) while the second idea is to generate it synthetically using an AO simulator, injecting experimental inputs (\cite{oberti2006large}, \cite{kolb2012whatcan}, \cite{kolb2012calibration}). A summary of the classical procedure and of some of these new strategies is developed in Section \ref{CalibrationAoSystem}.

The community seems to be converging towards the second approach, working on synthetic models of the AO systems in which are injected some mis-registrations parameters to fit with the alignment of the real system. This approach is especially suited for systems that require a frequent update of the calibration with a large number of actuators. The Pseudo Synthetic Interaction Matrix (PSIM) is currently the baseline for the AOF instruments in Ground Layer Adaptive Optics (GLAO) and Laser Tomography Adaptive Optics (LTAO) modes working with a Shack-Hartmann WFS (\cite{hartmann1900bermerkungen}, \cite{shack1971production}). 

It is however still to be investigated in the case of the LBT that operates in Single Conjugate Adaptive Optics (SCAO) mode, working with a pyramid WFS (\cite{ragazzoni1996pupil}) and Natural Guide Stars (NGS). The pyramid WFS specificities could add constraints for the calibration. This WFS provides a gain in sensitivity with respect to the Shack Hartmann (\cite{ragazzoni1999sensitivity}) but can be complex to model with a modal linearity and sensitivity (\cite{esposito2001pyramid}, \cite{verinaud2004nature}, \cite{fauvarque2017general}). The response of the sensor is also seeing dependent (\cite{ragazzoni1999sensitivity}) as the PSF shape on the top of the pyramid will depend on the seeing conditions. Some work has also shown that tracking the modal gains during the operations will improve the closed loop performance (\cite{korkiakoski2008improving}, \cite{esposito2012natural}, \cite{esposito2015NCPA}, \cite{bond2018optimized}, \cite{deo2018modal}). Considering that all of the first light instruments of the ELT will include a pyramid WFS in their design (\cite{neichel2016adaptive}, \cite{clenet2016joint}, \cite{brandl2016status}) it is necessary to identify the key-elements and the accuracy requirements to reproduce the behaviour of an AO system with pyramid WFS with the overall goal to generate calibration data that can be used on a real system. 

After a short introduction of the classical and new calibration methods in the context of the adaptive telescopes (section \ref{CalibrationAoSystem}), this paper will introduce  the development of a synthetic model, reproducing the FLAO-LBT systems (\cite{esposito2010first}), focusing on the model definition and sensitivity (section \ref{ModelingLBTAOsystems}). Section \ref{SignalsModelAlignment} details the adjustment procedure for the mis-registrations parameters that have been thoroughly verified in simulation. Section \ref{validationLBT} presents the results of day-time validation at the LBT.

\section{AO Calibration of an Adaptive Telescope}
\label{CalibrationAoSystem}
This section aims to present the calibration procedure for a classical AO system and provides a short summary of the different calibration strategies in the frame of the future Extremely Large Telescopes. 

\subsection{General case}
Mathematically, the interaction matrix of an AO system is the transfer matrix between the DM and the WFS space. Following the notations introduced in \cite{meimon2015optimized}, the interaction matrix $D$ of an AO system is:
\begin{equation}\label{eq:1}
D=M_{WFS}.M_{DM}
\end{equation}
where $M_{DM}$ is the conversion matrix between the DM commands and the optical phase deformations and $M_{WFS}$ the corresponding WFS measurement matrix. This matrix is then inverted to provide the reconstructor $R$ that is used in closed loop. 
\begin{equation}
R=D^\dag
\end{equation}
The most common inversion method consists of using a Truncated Singular Values Decomposition (SVD), filtering the modes badly seen by the WFS for stability (\cite{boyer1990adaptive} but using a Generalised SVD adding priors on the noise and turbulence statistics provides a gain in the reconstruction (\cite{wallner1983optimal}, \cite{fusco2001optimal}, \cite{gilles2005closed}). 

The measurement of the interaction matrix is achieved by recording the WFS signals $Y$ corresponding to a given set of actuation pattern $U$. These measurements are impacted by two sources of noise, the WFS detector noise $W$ and the local turbulence $\Phi$ during the measurements. 
\begin{equation} \label{eq:2}
Y=DU+W+M_{WFS}.\Phi 
\end{equation}
Multiplying (\ref{eq:2}) by the pseudo inverse of the calibration pattern $U^\dag$ provides the estimation of the interaction matrix ${\widehat{D}}$.
\begin{equation} \label{eq:3}
\widehat{D} =Y.U^\dag
\end{equation}
The calibration pattern is usually chosen to be full rank (for instance a zonal actuation or a modal basis such as DM stiffness modes or Kharunen-Lo\`eve modes) to calibrate all the degrees of freedom of the DM, providing $U.U^\dag=Id$. That way, we can define the calibration error $\Delta D$:
\begin{equation}\label{eq:4}
\Delta D= D-\widehat{D}=-(W+M_{WFS}\Phi).U^\dag
\end{equation}

This equation consists of two independent terms, $W$ the noise related to the WFS detector and  $M_{WFS}.\Phi$ the noise related to the turbulence, both multiplied by the pseudo inverse of the calibration pattern $U$.

In the optimal case, both noise contribution can be minimized using a bright artificial calibration source and reducing $\Phi$ to the local turbulence of the optical bench. In that case, the choice of calibration patterns becomes arbitrary as a good SNR is easily achievable. However, if these contributions become non negligible, and if the number of actuators becomes important, it is necessary to optimise the choice of calibration pattern to maximize the SNR and speed up the measurement. Such optimisation have already been investigated by \cite{oberti2004calibration}, \cite{kasper2004fast} and \cite{meimon2015optimized} showing that using system modes or Hadamard patterns to maximize the signal provides a consistent gain in calibration time. 

In any case, quantifying the quality of an interaction matrix is still an open question. Some metrics have been investigated in \cite{meimon2015optimized} and \cite{oberti2004calibration} but no clear criteria have been identified yet. The only solution to validate the accuracy of an interaction matrix is then to try using it to close the loop of the AO system.

\subsection{On-sky calibration}
With no calibration source available, an alternative method consists of measuring the interaction matrix on-sky. This strategy faces one major challenge: the impact of the turbulence on the WFS measurements. In equation \ref{eq:2}, the term $M_{WFS}.\Phi$ becomes indeed dominant. Actuating a calibration pattern using a classical push-pull would require such a large amplitude to extract it from the turbulence noise that it will completely saturate the WFS. 

The problem becomes then pretty classical: one wants to retrieve a low signal in a noisy (turbulence noise) and variable environment (seeing variation), with a constraint of not saturating the WFS nor the DM stroke. 

Taking the two extreme solutions, one could either average the turbulence using a long integration, or freeze it using a fast push-pull measurement. An alternative approach inspired from the telecommunication domain, consists of modulating signals with a low amplitude on the DM (to minimize the impact on science) but large enough to be detectable by the WFS. Using an orthogonal basis, in both temporal and spatial domain, of multiplexed signals, one can retrieve their signatures in the WFS space using a simple demodulation process (\cite{esposito2006onSkyIM}). This approach has been validated on several systems, at the VLT with a curvature sensor by \cite{oberti2006large}, with a high order pyramid WFS at the LBT by \cite{pinna2012first} (operating the demodulation in the Fourier space) and with a Shack Hartmann on the CANARY Facility (\cite{myers2008canary}) by \cite{kellerer2012deformable} (demodulating in the direct space).

The analysis of the results has shown that an accurate knowledge of the temporal behaviour of the system is required (rejection transfer function of the AO loop, temporal delay of the loop) and a trade-off for the modulation parameters (frequency, amplitude, sampling, multiplexing) has to be done to perform an efficient calibration. If these methods are applied during the observation, the impact on science is still to be investigated. In addition, to overcome the constraint of WFS saturation, the signals modulation has to be operated in closed-loop (at least partial) using a first interaction matrix, that could be synthetic.  

This demonstrated the feasibility of the on-sky approach as a potential alternative strategy but to our knowledge no operating system is currently using it as its baseline for its AO calibration. Such a calibration requires indeed a longer time than a classical measurement with fibre (with a lower SNR). Moreover, in the case of a Pyramid WFS, if the on-sky calibration takes too long, the WFS response may evolve during the measurements, as it depends on the seeing conditions, and bias the interaction matrix measurement. Therefore, a full on sky calibration does not seem suited for a system with a large number of degrees freedom, especially with pyramid WFS. In the case of adaptive telescopes, potential non linearities and complex internal behaviour (evolution or instability of the DM/WFS registration) may also appear during the operation.
These techniques could however be used to retrieve only a few signals on-sky to identify key mis-registrations parameters that are then injected in a synthetic model. This pseudo-synthetic approach seems then to be a better strategy to overcome such constraints.

\subsection{New pseudo-synthetic calibration}
 \begin{figure*}
 \begin{center}
	 \includegraphics[scale=1]{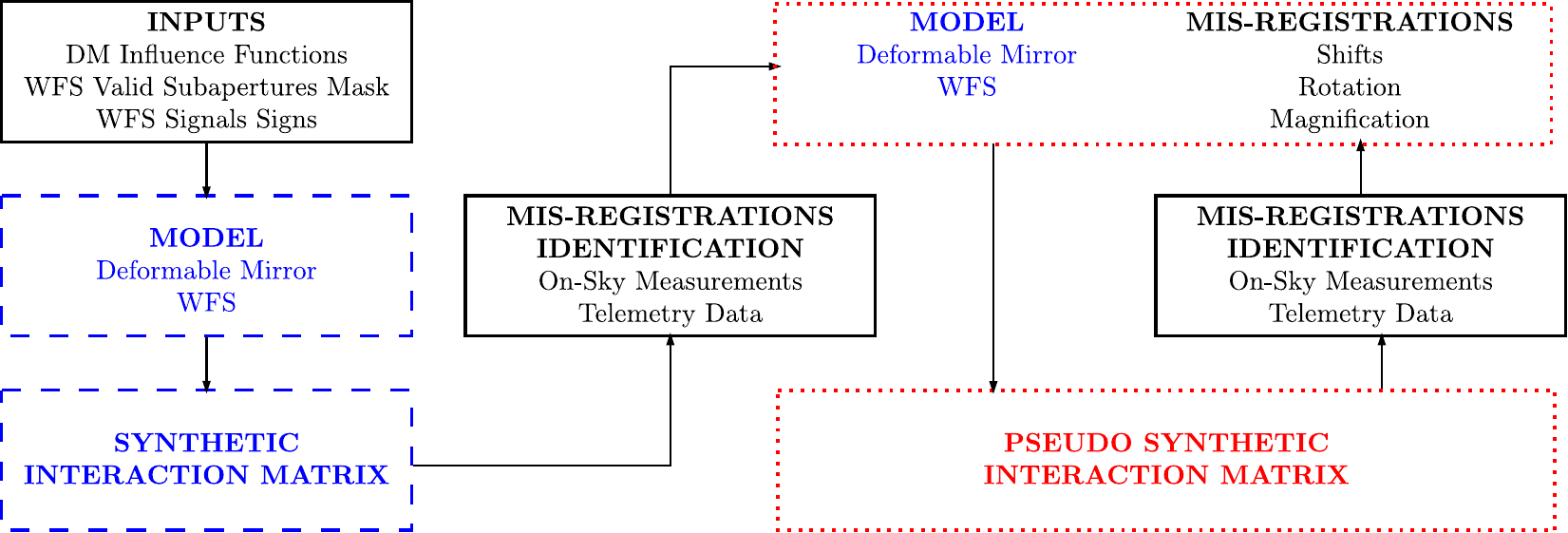} 
         \caption{Pseudo Synthetic Calibration: Experimental inputs (solid black lines) are injected into the synthetic model (dashed blue lines) to reproduce the registration of the real system. The pseudo-synthetic interaction matrix (dotted red lines) is then regularly updated during the operation tracking the mis-registrations parameters.}
          \label{fig:PSIM}
       \end{center}
\end{figure*}

The pseudo-synthetic calibration has been identified as the most promising calibration strategy for the future Extremely Large Telescopes: mis-registrations parameters extracted from experimental inputs\footnote{That we will call reference signals in the following of the paper.} are injected into a synthetic model of the AO systems. The principle is summarised in Fig. \ref{fig:PSIM}.

This approach provides a noise free interaction matrix, fast to compute and thus, easy to update. The good quality of the calibration relies only on two key ingredients: the ability to model accurately the WFS and the DM and the accuracy of the mis-registrations parameters. Theoretically, if all the experimental effects are perfectly reproduced, a Pseudo Synthetic interaction matrix should then provide a better calibration than a noisy experimental one. 

Concerning the first ingredient, the state of the art in terms of AO modeling has shown on various systems (e.g. LBT or VLT) that an accurate calibration can be generated from a synthetic model (\cite{oberti2006large}, \cite{kolb2012calibration}, \cite{pinna2012first}). The modeling of a Shack Hartmann seems to be one step ahead with respect to the Pyramid WFS case as extensive studies related to the sensitivity are available( see \cite{kolb2012calibration}). In the case of a Pyramid WFS, the feasability was demonstrated in \cite{pinna2012first} but no analysis of sensitivity was achieved so far. In the following of this paper, we will try to address this point. 

The second ingredient is related to the estimation of the mis-registrations parameters. The nature of the reference signals required to align the model is still open to discussion. The first approach is non-invasive and consists of using telemetry data (WFS slopes and DM commands) to retrieve the corresponding mis-registrations parameters. At the AOF for instance, the idea consists of using closed-loop data to estimate a noisy interaction matrix and project it on sensitivity matrices to identify these parameters (see \cite{kolb2012calibration} and \cite{bechet1a2011identification}).

The second approach is invasive and consists of dithering signals on the DM to retrieve their signature in the WFS space (\cite{chiuso2010dynamic}). In \cite{neichel2012identification}, a strategy based on a Levemberg-Marquardt (\cite{marquardt1963algorithm}) type algorithm is presented in the frame of tomographic AO systems but requires an experimental interaction matrix as a reference, that could for instance be measured on-sky.
The difficulty here is to find the minimum number of signals that would be characteristic of each type of mis-registrations, independently from the others. Since the parameters are strongly correlated, the solution consists here in using an iterative procedure to identify each parameter.

Optimising the accuracy of a pseudo-synthetic interaction matrix is another question. The matrix generated is indeed noise-free and equation \ref{eq:4} takes another form as the only source of error comes from the model: both $M_{DM}$ and $M_{WFS}$ become $\widehat{M_{DM}}$ and $\widehat{M_{WFS}}$. So far, the only way consists of validating the model against experimental inputs.

\section{Modelling of the LBT AO systems for a Pseudo Synthetic Interaction Matrix Calibration}
\label{ModelingLBTAOsystems}
\subsection{Model definition}
\begin{figure*}
 \begin{center}
	 \includegraphics[scale=1]{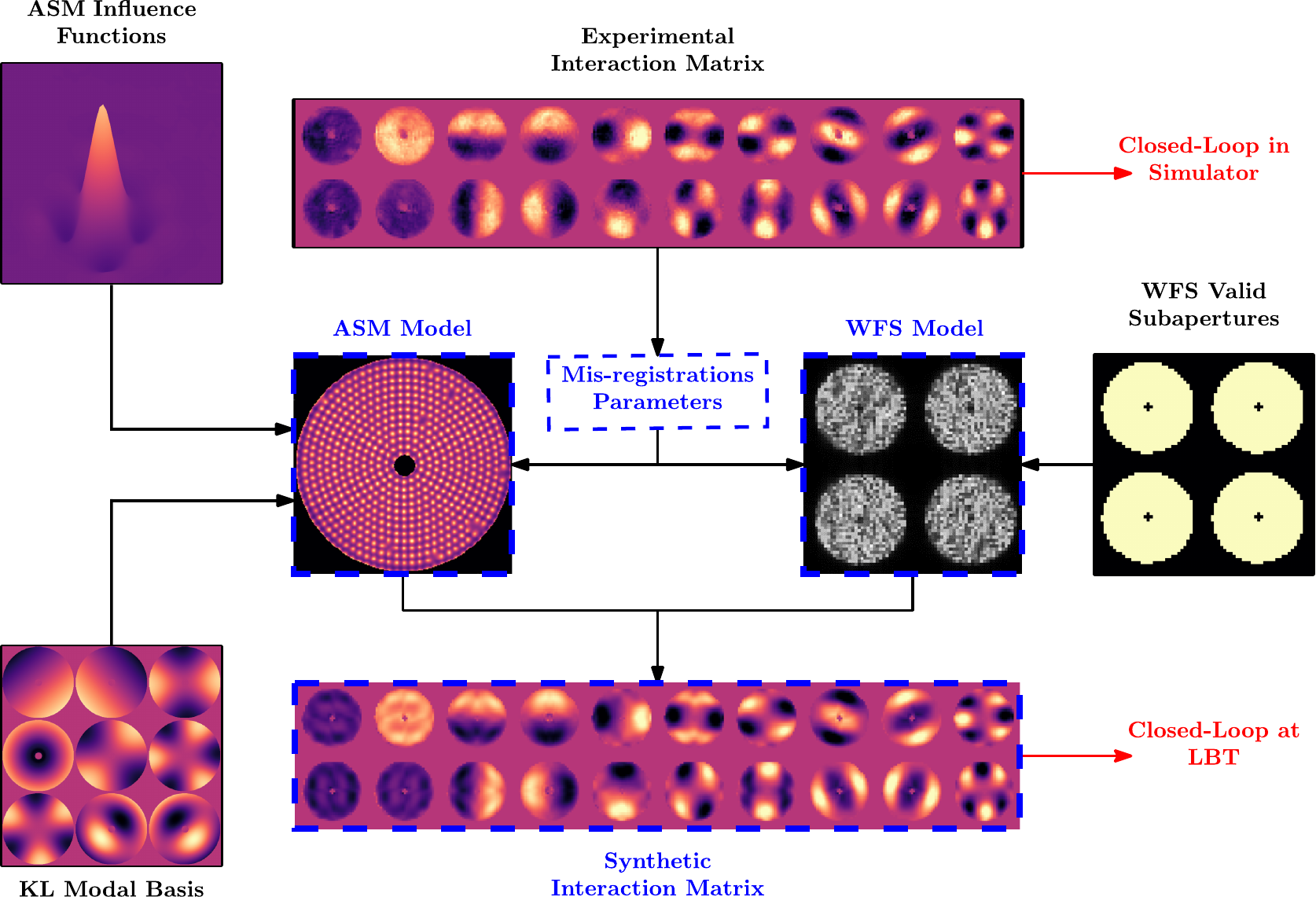} 
     \caption{Summary of the model definition. Experimental inputs (solid black lines) are injected into the synthetic model of the AO systems (dashed blue lines). The model mis-registrations parameters are first calibrated so that the experimental interaction matrix can be used to closed the loop in the simulator. The synthetic interaction matrix can then be tested on the real system.}
          \label{fig:LBT_layout}
       \end{center}
\end{figure*}

We reproduce the FLAO systems (\cite{esposito2010first}) at the LBT in the end to end simulator OOMAO (\cite{conan2014object}) with the overall goal to generate an interaction matrix that can be used on the real system. 
It required a fine tuning of the two key elements of the model, ASM and Pyramid WFS, using experimental inputs from the telescope to take in consideration all the features of the existing systems. A summary of the model definition is given in Fig. \ref{fig:LBT_layout}.

The \textbf{Adaptive Secondary Mirror} model is composed of 663 valid actuators arranged in circular concentric rings with a 30 cm radial pitch projected on-sky. The ASM Influence Functions measured with interferometer are input to the model. The modal commands matrix from the telescope to produce the 594 modes on the ASM is used in the simulator. The modal basis consists of Kharunen-Lo\`eve modes, computed by diagonalising the covariance matrix of the turbulence and re-orthogonalised in the DM space (\cite{esposito2010laboratory}). The amplitude of the modes used for the interaction matrix computation is low enough to prevent non-linearity effects.
As in the real system, the ASM is the stop aperture of the optical system. The arm of the telescope is not taken into consideration as it is done on site for the interaction matrix measurement procedure, achieved using a retro-reflector during day-time (\cite{esposito2010laboratory}). The four spiders of the retro-reflector are not considered either.

The \textbf{Pyramid WFS} diffractive model is based on the FLAO pyramid WFS with 30 by 30 subapertures (1 pixel per subaperture) and 36 pixels separating each pupil image centres. The WFS model is a perfect single pyramid\footnote{ The FLAO pyramid WFS is a double pyramid to avoid chromatic dispersion (\cite{esposito2010laboratory}) but the WFS model is achromatic as it operates at a single wavelength.} with no scratches or faces mis-alignment and operates at 750 nm, the central wavelength of the FLAO WFS. The FLAO valid subapertures mask is input to the model, selecting its position to maximize the amount of light in the WFS model pupils.
The WFS model is using the optimal Tip/Tilt modulation\footnote{No tuning of the Tip/Tilt modulation was achieved so far.} of 3 $\lambda/D$ considered at the telescope during the observations in high order mode. To reproduce the exact LBT configuration for the interaction matrix measurement, the WFS signals are normalised with a factor 2 to be in line with the double pass procedure using a retro-reflector (\cite{esposito2010laboratory}). 

\begin{figure}
 \begin{center}
\includegraphics[scale=1]{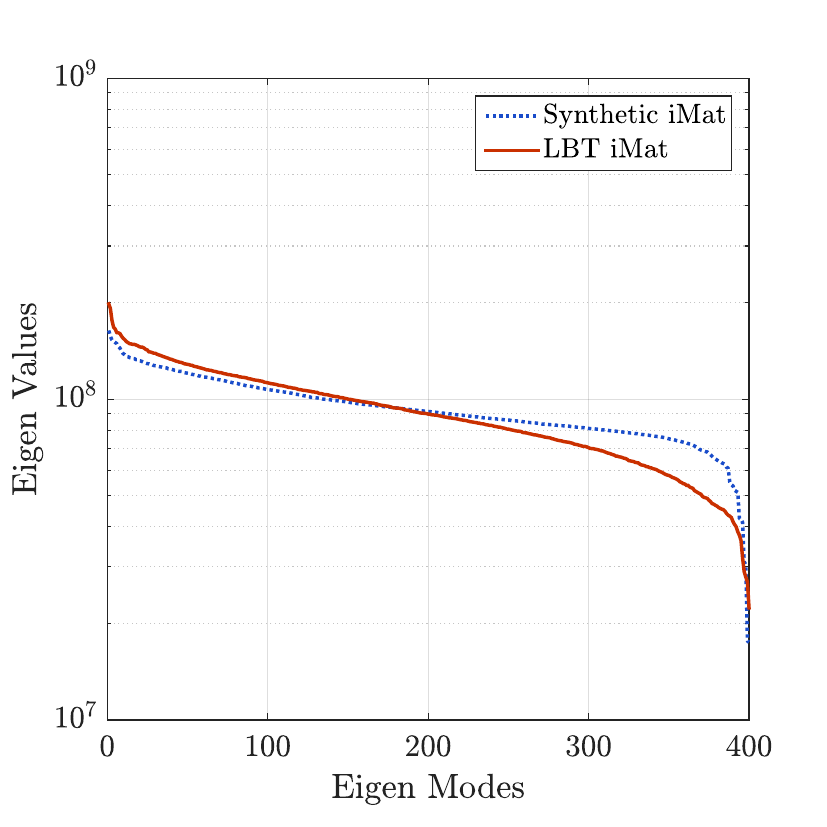}
 \caption{Comparison of the interaction matrices eigen Values distributions between the experimental and synthetic cases. The synthetic interaction matrix has a lower conditioning number (8.1 vs 8.85).}
  \label{fig:eigenValues}
 \end{center}
 \end{figure}

The computation of the synthetic interaction matrix is achieved in a noise-free environment. This feature is well illustrated on the interaction matrix eigen values spectra (Fig. \ref{fig:eigenValues}). In the synthetic case, the distribution of eigen values is flatter and the knee of the curve occurs for a larger eigen mode number. Here, note that the eigen modes will be slightly different as we still have residual differences between the two interaction matrices. However, the synthetic interaction matrix is better conditioned suggesting that it would be easier to control more modes using a synthetic interaction matrix than an experimental one, taking profit from the infinite SNR of the synthetic WFS signals. These considerations are however only valid if the registration of the real system is well reproduced in the simulator.

\begin{figure}
 \begin{center}
\includegraphics[scale=1]{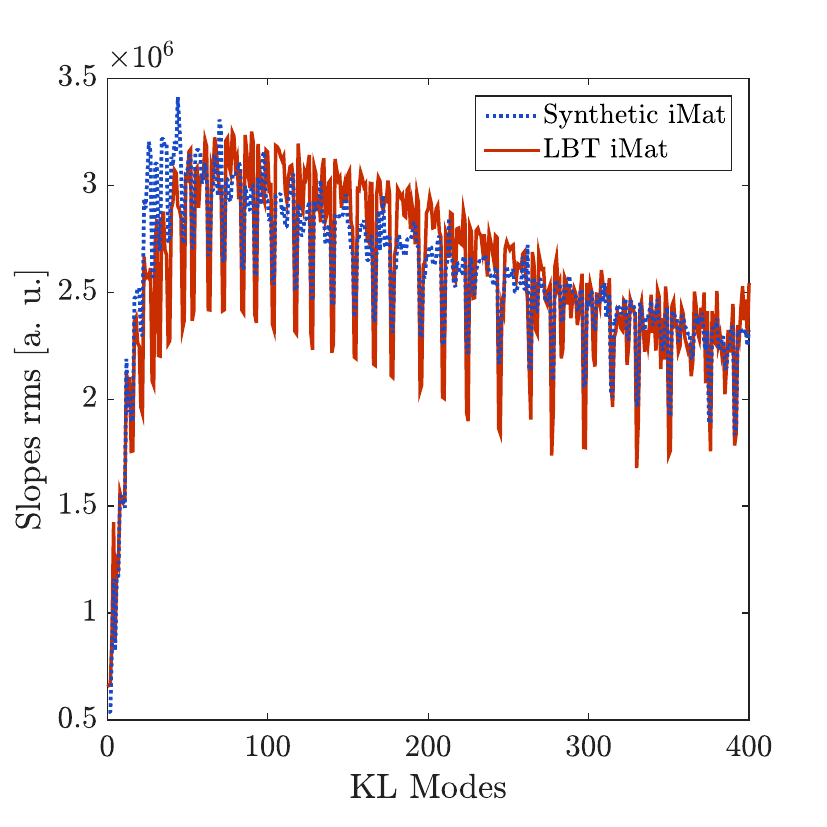}
 \caption{Comparison of slopes RMS plots. Both WFS have the same sensitivity to the KL modes.}
  \label{fig:slopesRMS}
 \end{center}
 \end{figure}

The Pyramid sensitivity to the modal basis depends on the Tip/Tilt modulation used for the WFS. A rough estimation of the WFS sensitivity is obtained by considering the WFS slopes Root Mean Square (RMS). The comparison between the model and the experimental WFS is provided in Fig. \ref{fig:slopesRMS}. Both sensitivity plots follow the same tendency which show that the model reproduces well the real system with the same parameter values.
 
 \subsection{Sensitivity to a mis-registration}

The accuracy of the model mis-registrations parameters is a key ingredient for high order AO systems. In practice, with a Shack Hartmann WFS, the common rule is not to exceed an error of 10 \% of a subaperture (\cite{bechet1a2011identification}). This result has to be taken carefully as the sensitivity to a mis-registration will depend on the DM geometry and on the number and type of modes controlled in the reconstructor. 

\begin{figure}
 \begin{center}
\includegraphics[scale=1]{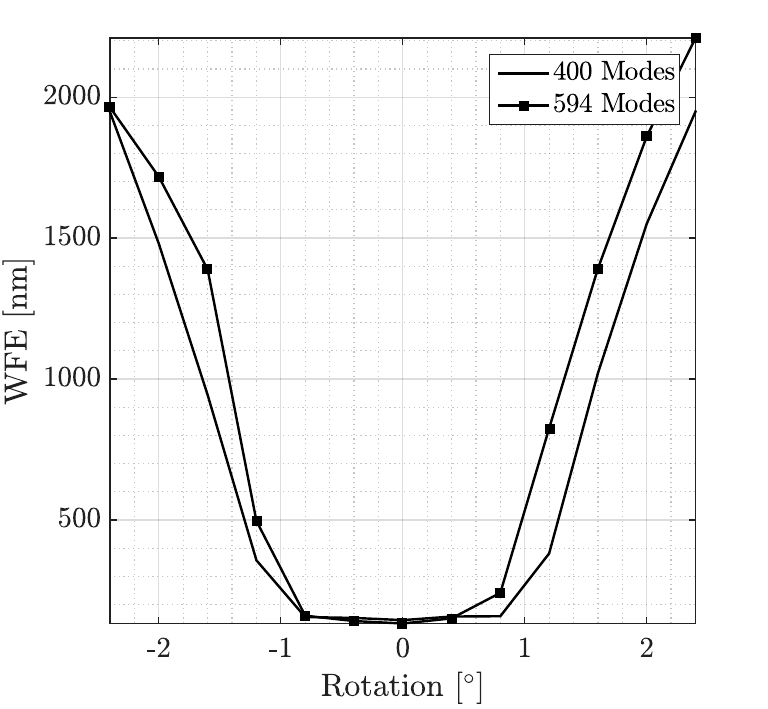}
\includegraphics[scale=1]{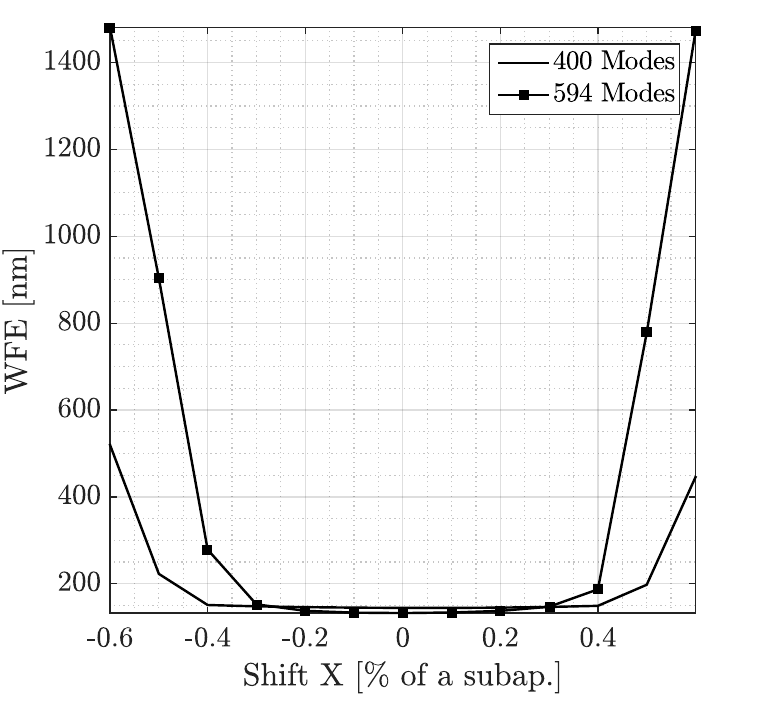}
\includegraphics[scale=1]{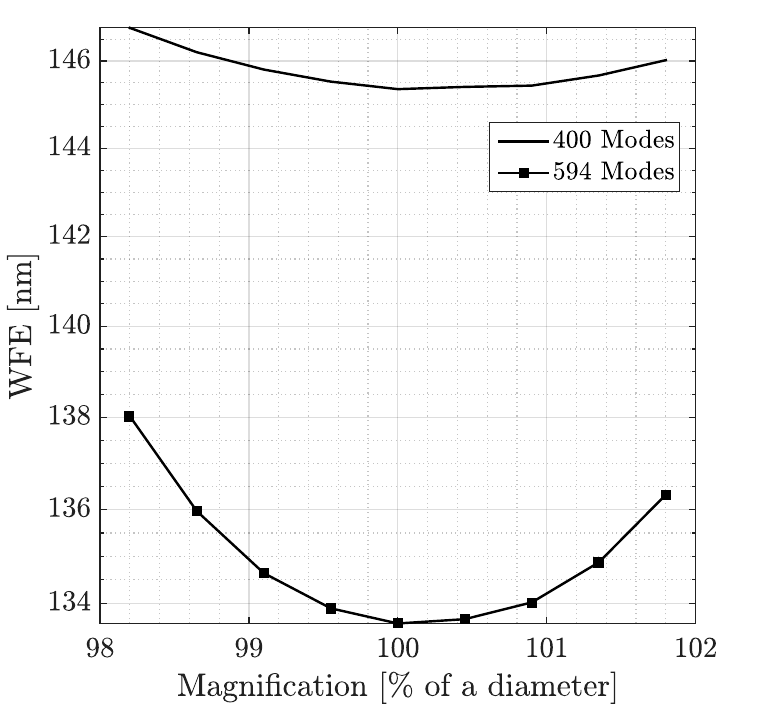}
 \caption{Impact of a mis-registration on the performance (Wave-Front RMS) for a rotation (Top), a shift in X (Middle) and a magnification (Bottom). Results are given for two configurations, controlling 400 and 594 modes.}
  \label{fig:misRegistrationPerformance}
 \end{center}
 \end{figure}

Using our model of Pyramid WFS and ASM with circular geometry, we could simulate the impact of each type of mis-registrations on the performance for a seeing of 1" in the visible. The corresponding Wave-Front Error (WFE) as a function of each type of mis-registrations, controlling 400 and 594 modes in the reconstructors is given Fig. \ref{fig:misRegistrationPerformance}. In the case of the LBT, a rotation of $1^\circ$ corresponds to a shift of around 25 \% of a subaperture on the border of the pupil.

\begin{figure}
 \begin{center}
\includegraphics[scale=1]{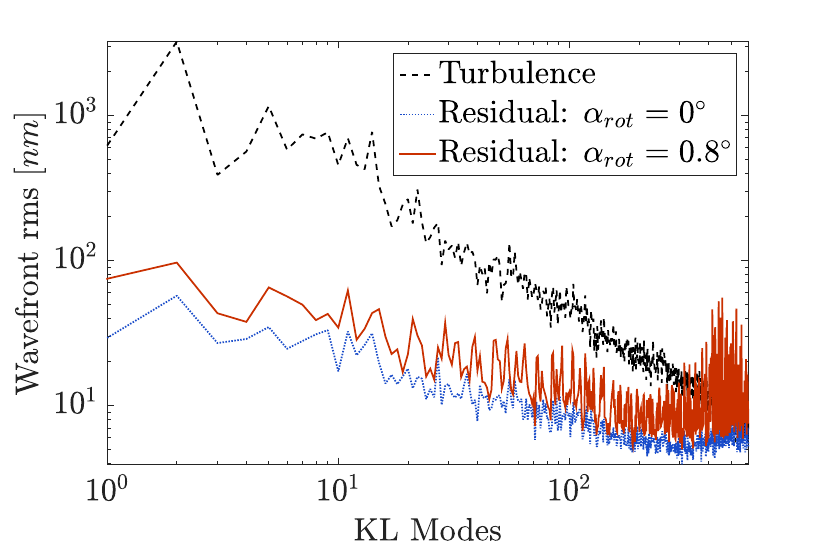}
\includegraphics[scale=1]{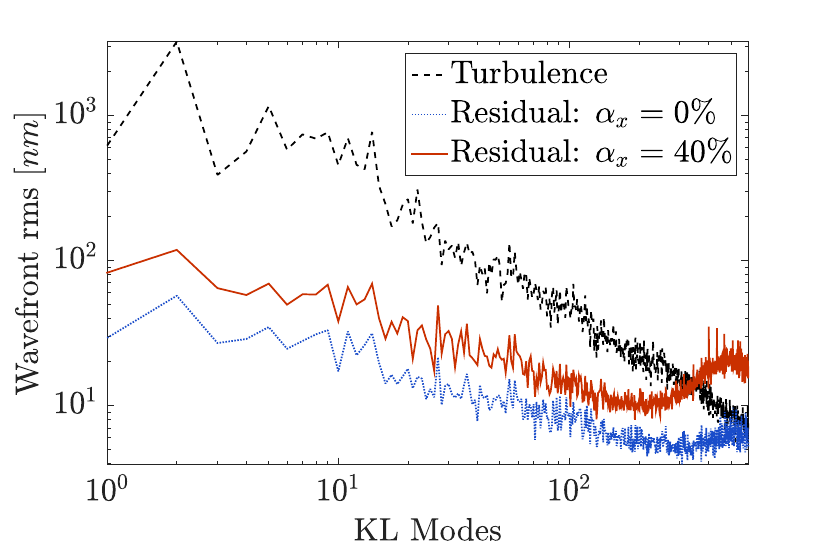}
 \caption{Modal PSD for the nominal and a mis-registrated case in the case of a rotation (Top) and a shift in X (Bottom)}
  \label{fig:PSDvsMisRegistration}
 \end{center}
 \end{figure}

More details are provided Fig. \ref{fig:PSDvsMisRegistration} displaying the modal PSD corresponding to the first mis-registration value after the drop of performance when controlling 594 modes ($0.8^\circ$ for the rotation and 40 \% for the shift). It is clearly visible that the high order modes are the most impacted by the mis-registrations and get amplified even to higher values than the incoming turbulence, confirming that high order AO systems are more sensitive to mis-registrations. 

These results show that the most critical mis-registrations parameters, the shifts and the rotation should be accurately identified (with less than 10 \% of a subaperture for the shifts and less than 0.1$^\circ$ for the rotation which corresponds to a shift of 2.5 \% on the border of the pupil) while the impact of a magnification becomes significant when controlling a high number of modes only. 

\section{ASM/WFS mis-registrations adjustment procedure}
\label{SignalsModelAlignment}
This section describes the procedures to finely tune the model mis-registrations parameters and provide a functional interaction matrix for the telescope. We consider here only the fine tuning and rely on a first rough estimation of the parameters (+/- 1 subaperture shift and +/- 2$^\circ$ of rotation) that could be done using flux considerations for instance (\cite{kolb2012whatcan}).

The tuning of the model requires to define a metric to quantify the error between the model and the real system (\ref{minimizationCriteria}). In our case, we use reference WFS signals that correspond to the measurement of 400 KL modes (the current interaction matrix used at the telescope). We then play on the model mis-registrations parameters to generate the corresponding synthetic WFS measurements and use an iterative procedure to minimize the error with the reference.

\subsection{Mis-registrations Parameters}
\label{misRegistrationParameters}
The model has 4 \textbf{mis-registrations parameters}, the shifts $\alpha_x$ and $\alpha_y$, the rotation $\alpha_{rot}$ and the radial magnification $\alpha_{magn}$ as it appears to be symmetric.  
We define $\boldsymbol{\alpha}$ as the corresponding model mis-registrations parameter vector: 

\begin{equation}
\boldsymbol{\alpha} \triangleq \{\alpha_x,\alpha_y,\alpha_{rot},\alpha_{magn}\}
\end{equation}
In practice, the relative shift between WFS and ASM is applied by shifting the WFS pupils on the WFS detector pixel grid by tilting the pyramid model (providing a sub pixel sensitivity). Both rotation and magnification are applied on the ASM model, interpolating the influence functions. 

\subsection{Minimization Criteria}
\label{minimizationCriteria}
We define $\sigma_j$ as the Root Mean Square Error (RMSE) between the synthetic WFS measurement $Y^*(\boldsymbol{\alpha})$ in the configuration  $\boldsymbol{\alpha}$ and the reference WFS measurement $Y$ from a real interaction matrix of the mode $j$:
\begin{equation}
	\sigma_j=RMSE(Y_j,Y^*_j(\boldsymbol{\alpha}))=\sqrt{\frac{1}{N_S}\sum_{n=1}^{N_S} |Y_j-Y^*_j(\boldsymbol{\alpha})|^2 }
\end{equation}
where $N_S$ is the number of WFS slopes.
The criteria $\chi_N$ to minimize is the quadratic norm of $\sigma_N=\{\sigma_i\}_{_{i=1,2,...,N}}$ adjusting  $\boldsymbol{\alpha}$ and eventually considering different numbers of reference signals $N$:

\begin{equation}			
	\chi_N (\boldsymbol{\alpha}) = arg_{_{min}} {||\sigma_N||}^{2}
\end{equation}
The optimal value chosen for $\boldsymbol{\alpha}$ is the convergence value of the iterative minimization of $\chi_{400}$ using all the reference signals available.
\subsection{Model Adjustment Procedure}
 
The mis-registrations parameters of the model, especially the shifts and rotation, are strongly correlated in the WFS space. It is therefore necessary to achieve an iterative procedure to adjust correctly the mis-registrations parameters. The parameters are estimated one by one and the procedure is summarised in Table \ref{tab:example_table}.
A first estimation of each shift and rotation is achieved using a large step and then a second estimation using a smaller step. Applying a third estimation does not change the value of the estimated parameters therefore two steps for the iterative procedure are sufficient. The magnification is identified as a last step as its effect is less crucial (see Fig. \ref{fig:misRegistrationPerformance}).

The starting point $\boldsymbol{\alpha_0}$ is defined using the Tip and Tilt WFS signals to roughly estimate the starting rotation value as these modes do not have a circular symmetry.  
\begin{equation}
\boldsymbol{\alpha_0}=\{\alpha_{rot}=64.5^\circ , \alpha_x=0 \% , \alpha_y=0 \%, \alpha_{magn}= 100 \%\} 
\label{eq.8}
\end{equation}

\begin{figure}
 \begin{center}
\includegraphics[scale=1]{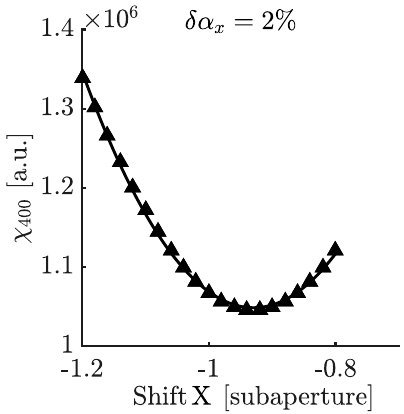}
\includegraphics[scale=1]{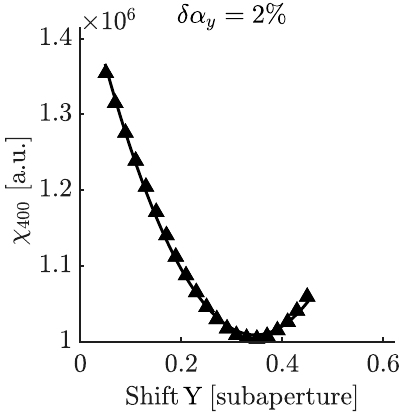}
\includegraphics[scale=1]{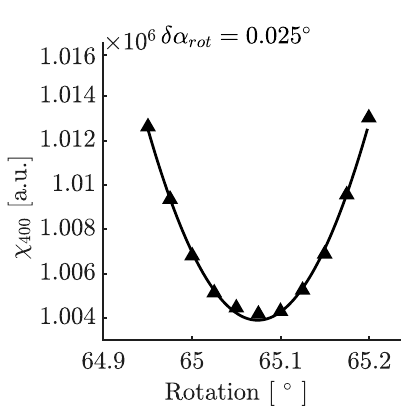}
\includegraphics[scale=1]{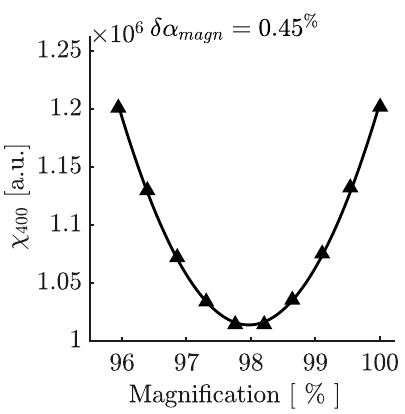}
 \caption{Last step of the minimization criteria $\chi_{400}$ for the shifts, rotation and magnification parameters using all the reference signals available. The solid line is a second order polynomial fit.}
  \label{fig:shiftsRotMagnEstimation}
 \end{center}
 \end{figure}

\begin{figure}
 \begin{center}
	 	 \includegraphics[scale=1]{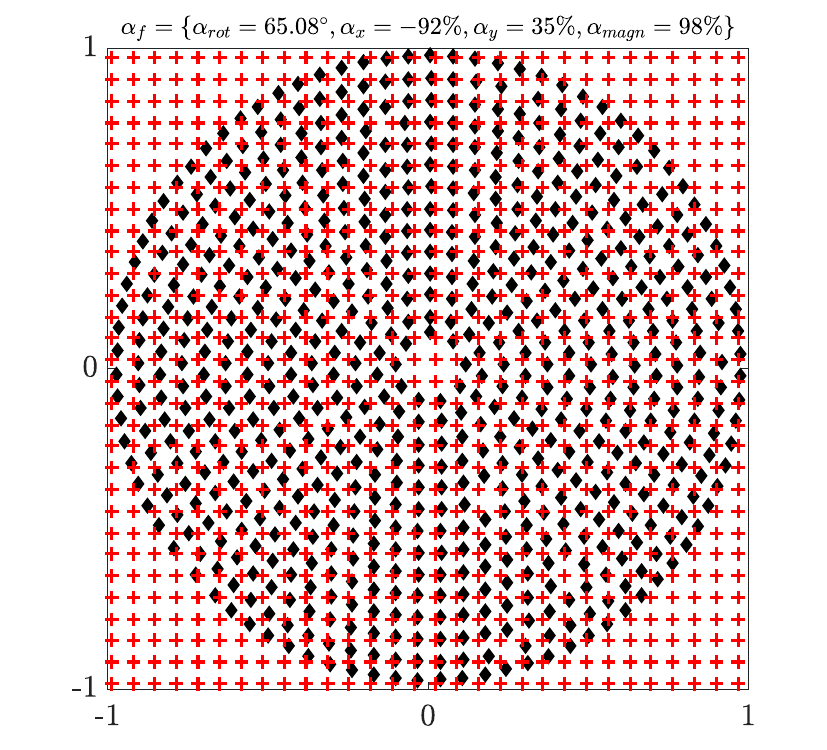} 
  \caption{Final configuration for the registration between the ASM actuators (diamonds) and the WFS subapertures (crosses) in normalised units.}
  \label{fig:AsmActuatorsWfsSubapertures}

 \end{center}
 \end{figure}

Fig. \ref{fig:shiftsRotMagnEstimation} gives the second step of the parameters estimation, showing a quadratic behaviour around each optimal value. The final value taken for the mis-registrations parameter $\boldsymbol{\alpha_f}$ is obtained from a second order polynomial fit of $\chi_{400}(\boldsymbol{\alpha})$ and is illsutrated in Fig. \ref{fig:AsmActuatorsWfsSubapertures}.
\begin{equation}
\boldsymbol{\alpha_f}=\{\alpha_{rot}=65.08^\circ , \alpha_x=-92 \% , \alpha_y=35 \%,\alpha_{magn}= 98 \%\} 
\end{equation}
\label{eq.9}
As a last step, we also tuned the modal gains of the pyramid model by playing on the amplitude of the KL modes during the interaction matrix computation and minimize the residual slopes RMS with the reference. The corresponding plots are given in Fig. \ref{fig:slopesRMS}.

\begin{table}
	\centering
	\caption{Model alignment iterative procedure. The values are in fraction of a subaperture (shift), in fraction of pupil diameter (magnification) and in degrees (rotation).}
	\label{tab:example_table}
\begin{tabular}{|c|c|c|c|}	
	\textbf{Step}&\textbf{$\alpha_i$}&\textbf{$\delta \alpha_i$}& \textbf{Value}\\
	\hline
	\hline
	1&$\alpha_{rot}$&0.125$^\circ$ &65.33$^\circ$\\
	\hline
	2 & $\alpha_x$&25 \% &-100 \%\\
	\hline
	3 & $\alpha_y$&25 \% &25 \%\\
	\hline
	4 &$\alpha_{rot}$&0.1$^\circ$ &65.08$^\circ$\\
	\hline
	5 &$\alpha_x$&2 \% &-92 \%\\
	\hline
	6 &$\alpha_y$&2 \% &35 \%\\	
	\hline
	7 &$\alpha_{rot}$&0.025$^\circ$ &65.08$^\circ$\\
	\hline
	8 &$\alpha_x$&2 \% &-92 \%\\
	\hline
	9 &$\alpha_y$&2 \% &35 \%\\	
	\hline
	10 &$\alpha_{magn}$&0.45 \% &98 \%\\
	\hline
	\end{tabular}
\end{table}

 
\subsection{Validation of the mis-registrations parameters adjustments in simulation}
\label{validationSimulator}
\begin{figure}
 \begin{center}
\includegraphics[scale=1]{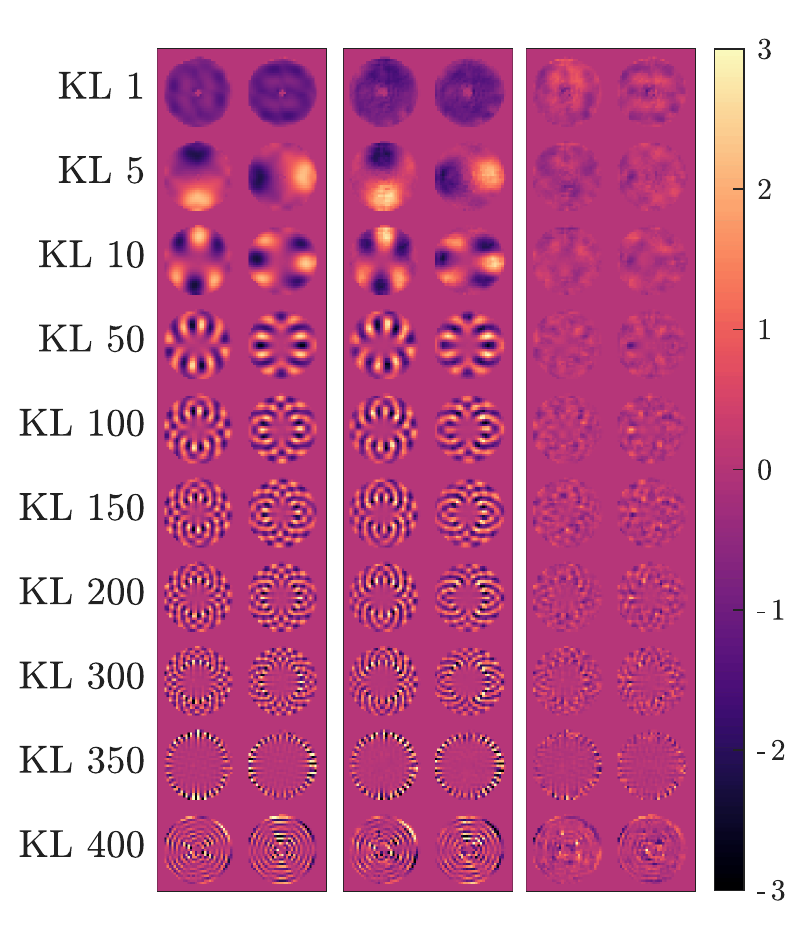}
 \caption{Comparison of pseudo synthetic (Left), LBT (Center) and residual (Right) normalised slopes maps after adjustment of the model. These slopes maps correspond to KL modes 1, 5, 10, 50, 100, 200, 300, 350 and 400.}
  \label{fig:slopesComparison}
 \end{center}
 \end{figure}

\begin{figure}
 \begin{center}
\includegraphics[scale=1]{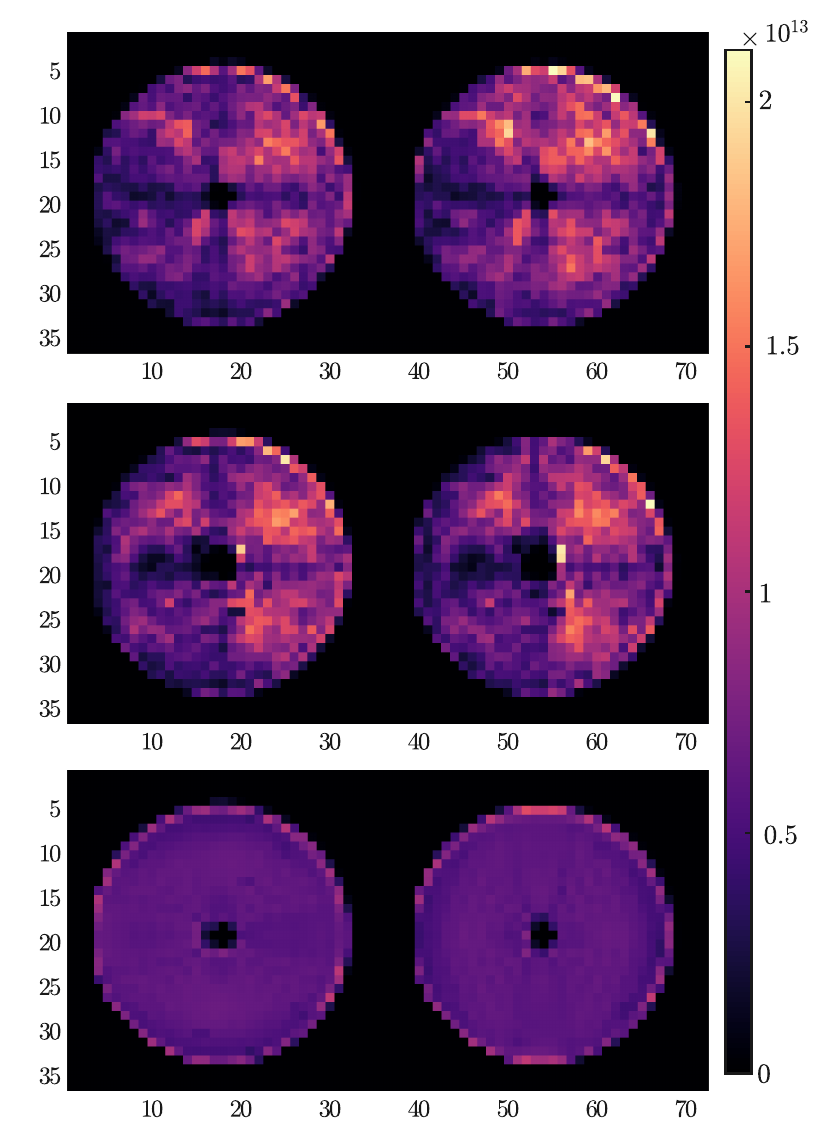}
 \caption{Slopes variance map [Sx Sy] corresponding to the 400 KL modes contained in the Interaction matrices of the LBT from 2016 (Top), 2017 (Middle) and from the simulator (Bottom). It is visible that some parts of the pupils gathered less signals than others, the bottom left part, the central obscuration and behind the spiders of the retro-reflector. These features were not considered in the model. }
  \label{fig:slopesVariance}
 \end{center}
 \end{figure}

The adjustment of the model was intensively validated in simulation. A comparison of a few slopes maps from both interaction matrices is given in Fig. \ref{fig:slopesComparison} and shows only negligible differences. It also shows that some features of the real measurements are missing from the model especially close to the central obscuration. These features seem to be purely experimental as evolving between different experimental interaction matrices (see Fig. \ref{fig:slopesVariance}). We did not consider these features in the model and their impact is still to be investigated.

\begin{figure}
 \begin{center}
\includegraphics[scale=1]{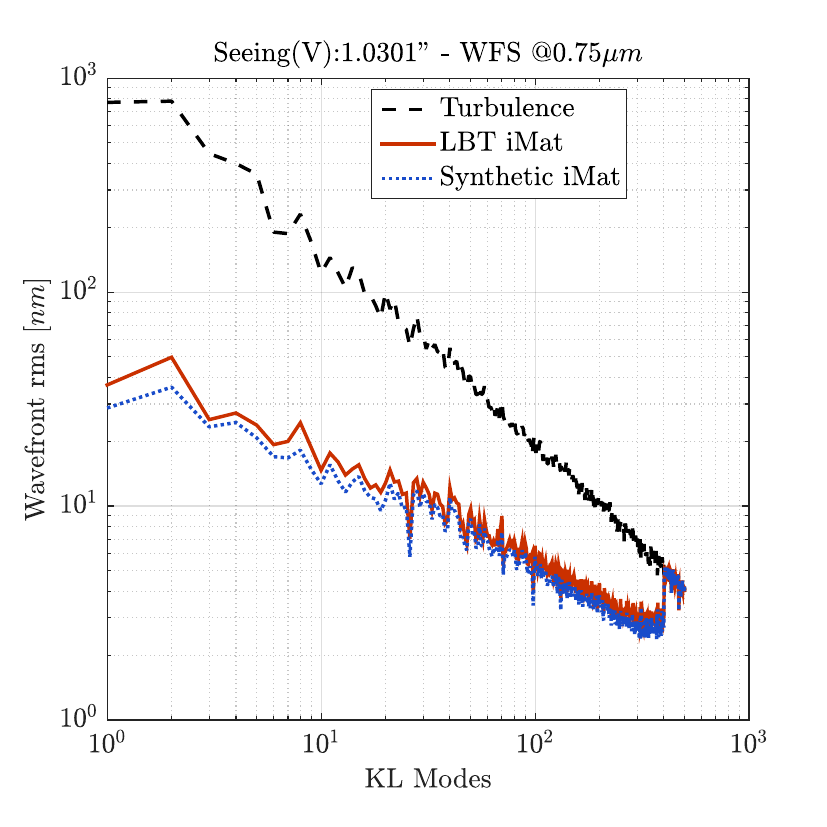} 
 \caption{Comparison of simulated closed loop Modal Power Spectral Density (PSD) using both reconstructors, synthetic and measured at the telescope.}
  \label{fig:comparisonCLperfSimulator}
 \end{center}
 \end{figure}

To provide a meaningful comparison, the first milestone was to use the interaction matrix measured at the telescope to close the loop of the simulated AO system. The comparison of simulated closed loop performance of both synthetic and measured interaction matrices provides an indication of the model accuracy. This is given in Fig. \ref{fig:comparisonCLperfSimulator} and as expected, the synthetic calibration provides a better correction (the corresponding calibration is the optimal calibration for the model) but overall, both reconstructors provide an equivalent correction with a residual variance of 128 nm in the synthetic case and 147 nm using the experimental reconstructor. The differences in residual variance come from the remaining model errors and/or calibration errors from the experimental data. 
To get rid of these residual differences in performance, some eventual upgrade of the model could be to take in consideration the imperfect illumination identified on the experimental interaction matrix to improve the accuracy of the model.

\begin{figure}
 \begin{center}
\includegraphics[scale=1]{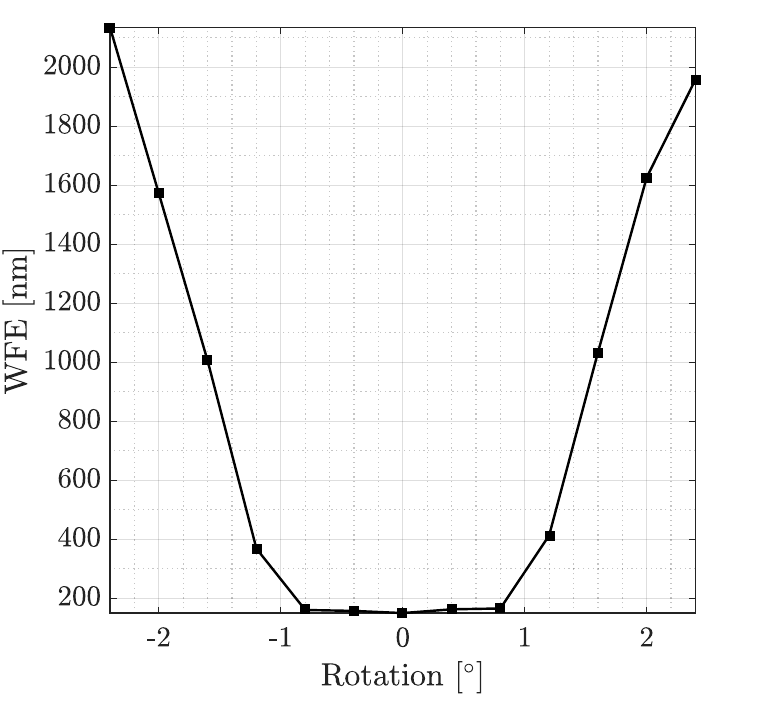} 
\includegraphics[scale=1]{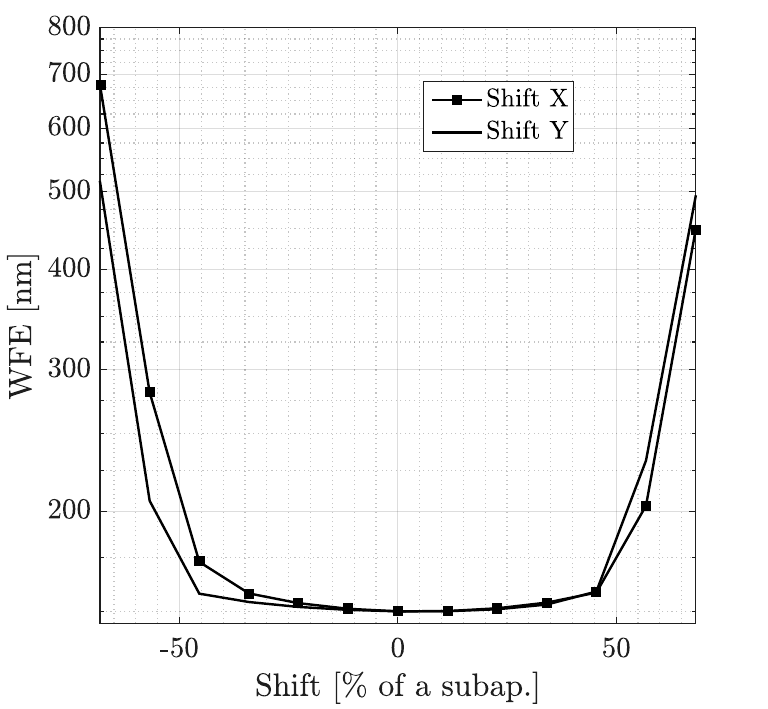} 
\includegraphics[scale=1]{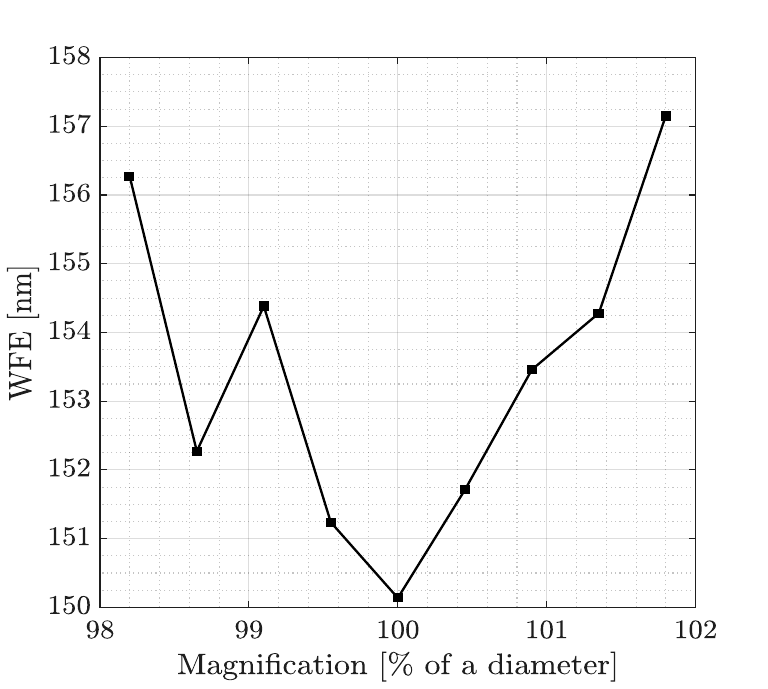} 

 \caption{Simulated closed loop performance using the experimental reconstructor in the same conditions as Fig. \ref{fig:misRegistrationPerformance}, playing around the optimal value of rotation (Top), both shifts (Middle) and magnification (Bottom) identified during the model adjustment procedure.} 
  \label{fig:comparisonCLperfExpvsMisRegSimulator}
 \end{center}
 \end{figure}

Moreover, Fig. \ref{fig:comparisonCLperfExpvsMisRegSimulator} gives the simulated closed loop performance using the experimental reconstructor and around the optimal mis-registration parameters value identified in section \ref{SignalsModelAlignment}. For each parameter, the optimal value provides the best AO performance, confirming the high accuracy of the mis-registrations parameters identification. We also retrieve the same sensitivity to the mis-registrations as in the synthetic case (see Fig. \ref{fig:misRegistrationPerformance}). 

\section{Experimental Validation}
\label{validationLBT}
The model was validated using day-time remote tests at the LBT. In that configuration, a star of magnitude 7 is simulated using an external fibre and a retro-reflector. A disturbance signal is then applied on the ASM to simulate a turbulence with a seeing of 1" in the visible. A summary of the calibration procedure of the LBT is developed in \cite{esposito2010laboratory}.

Before closing the loop, a tuning of the integrator gain is applied for the Tip/Tilt, the following modes up to 100 and for the remaining modes up to 400. The tuning of the gains consists of applying a ramp of gain values and select the one minimizing the RMSE of the residual slopes. 
That way, a different gain value will be used for each group of modes during the closed-loop operation. For our experimental tests, we applied this procedure to the experimental interaction matrix and used the corresponding values for our synthetic interaction matrix with no further tuning.

\begin{figure}
 \begin{center}
\includegraphics[scale=1]{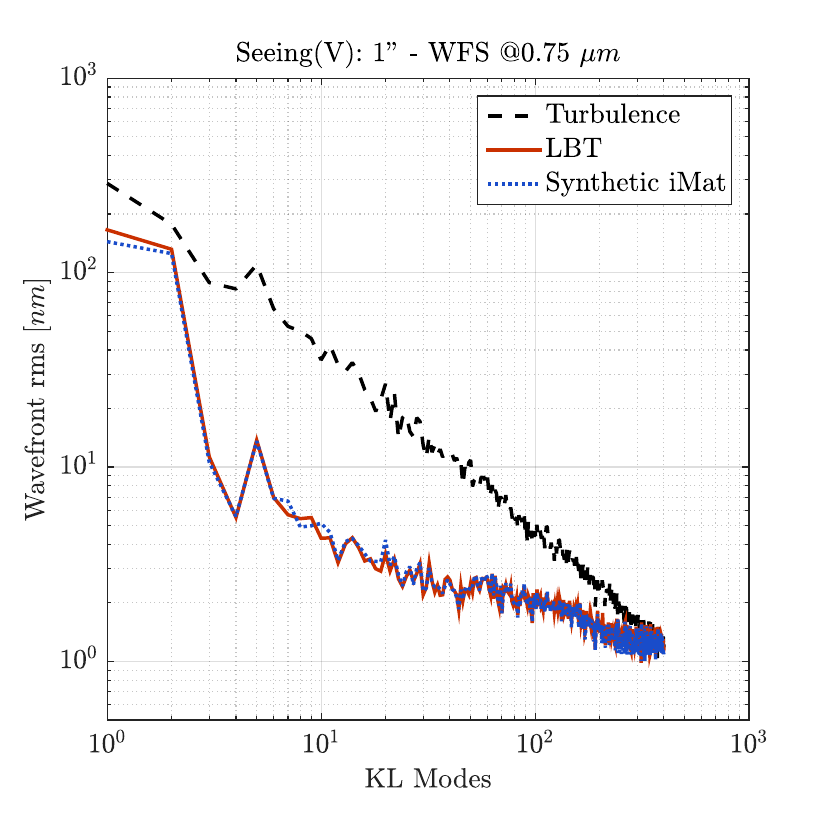}
 \caption{Comparison of modal PSD using both interaction matrices, Synthetic and Reference for a same disturbance applied on the ASM.}
 \label{fig:comparisonCLperfExperimental}
 \end{center}
 \end{figure}

Fig. \ref{fig:comparisonCLperfExperimental} provides the comparison of performance between the pseudo-synthetic and experimental interaction matrix using the mis-registrations values identified from the iterative procedure defined in \ref{SignalsModelAlignment}. Both reconstructors provide an equivalent correction and the details of the corresponding residuals are summarised in Table \ref{tab:recapExpLBT}, showing that the pseudo-synthetic interaction matrix provided a slightly better correction than the experimental one. We can notice on the modal PSD that some low order modes (mostly Tip/Tilt and Focus) have a significantly higher variance compared to what we could expect with a simulated turbulence and a bright star. This effect takes origin from the estimation of the residuals. These are computed from the DM positions which include the correction for the vibrations of the telescope. 
These modes should then not be considered for the performance comparison. 

\begin{figure}
 \begin{center}
\includegraphics[scale=1]{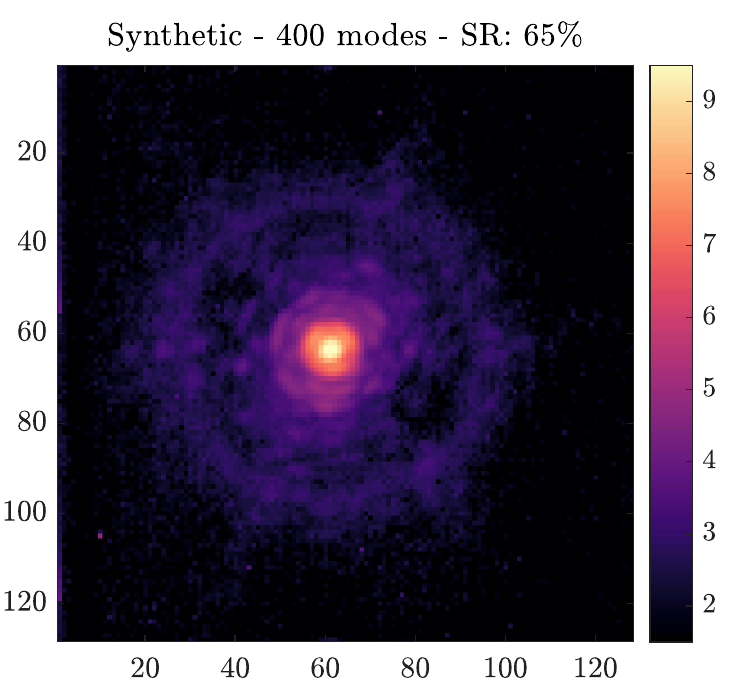}
\includegraphics[scale=1]{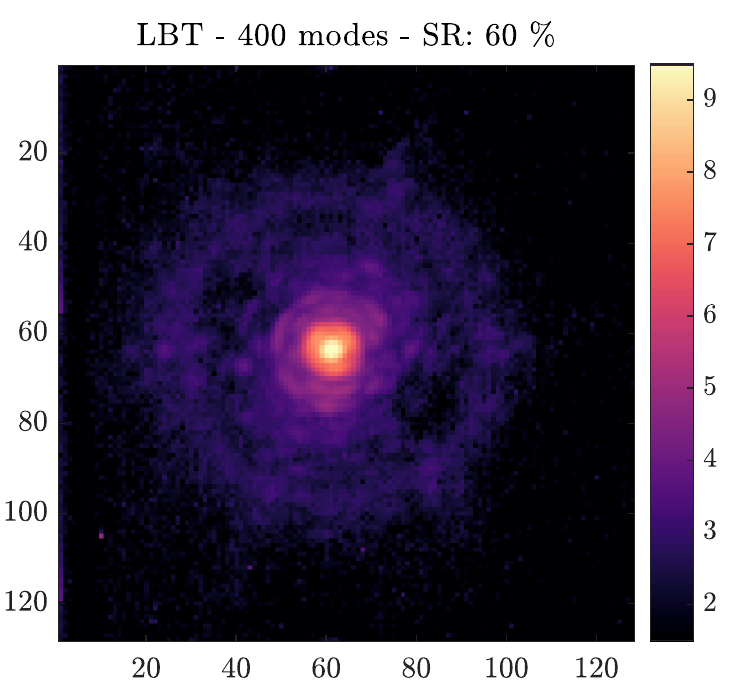}
\includegraphics[scale=1]{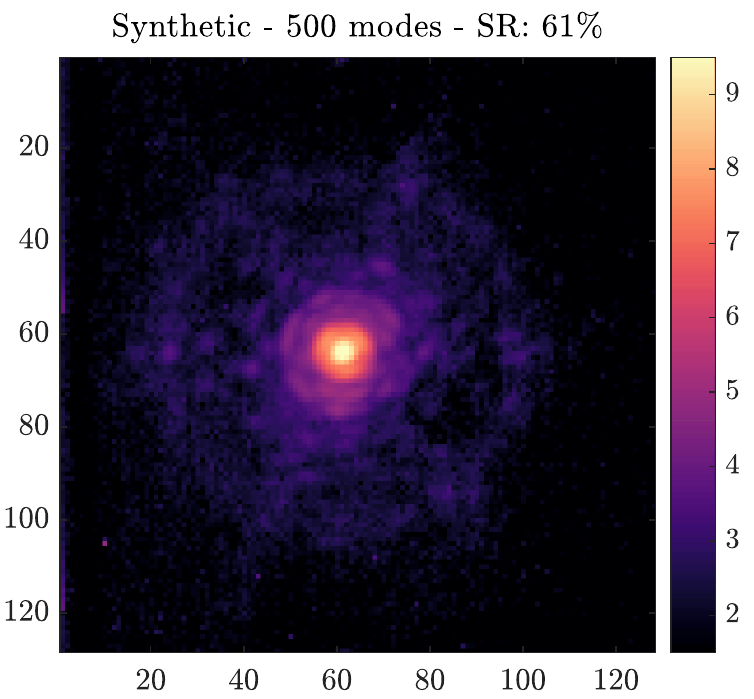}
 \caption{PSF in H band (log scale) controlling 400 and 500 modes with the synthetic and experimental interaction matrix. The Field of View of the instrument is 30". }
  \label{fig:PSF}
 \end{center}
 \end{figure}

In addition, we managed to push the synthetic interaction matrix to correct up to 500 modes and get stable closed-loop with high order correction visible in Fig. \ref{fig:PSF}. This validates the high accuracy of the mis-registrations parameters estimation while having WFS reference measurements for only 400 KL modes. The corresponding PSF from the instrument LUCI (\cite{heidt2018}) for 400 and 500 modes using the synthetic interaction matrix are also given in Fig. \ref{fig:PSF}. 
\begin{table}
	\centering
	\caption{Summary of the experimental closed loop performance for both pseudo synthetic and experimental interaction matrices.}
	\label{tab:recapExpLBT}

\begin{tabular}{|c|c|c|c|}	

	\textbf{Int. Mat.}&\textbf{WFE}&\textbf{TT rem.}&\textbf{50 KL rem.}\\
	\hline
	PSIM 400 Modes & 195 nm & 41.7 nm& 30.1 nm\\
	LBT 400 Modes & 216 nm & 42.1 nm& 30.8 nm\\
	PSIM 500 Modes & 228.3 nm &42.5 nm & 27.39 nm\\
	\hline
	\end{tabular}
\end{table}

\section{Conclusions}

We developed a synthetic model of the FLAO Pyramid WFS and Adaptive Secondary Mirror using experimental inputs from the telescope. We considered a perfect pyramid WFS and used measured influence functions for the ASM model. To identify the mis-registrations parameters, we defined an iterative procedure using an experimental interaction matrix as a reference to adjust the model mis-registrations parameters. 
The accuracy of these parameters has been thoroughly verified in simulation.

The interaction matrix generated from the model has been validated experimentally using day-time at the LBT. This demonstrates the feasibility of the pseudo synthetic calibration for high-order AO systems with pyramid WFS. Using the mis-registrations parameters identified from the experimental reference, no further tuning on-site was required to efficiently close the loop of the real system. It provided slightly better performance than the experimental one and we could control up to 500 modes with the synthetic reconstructor. 

Through this experimental validation, we have now a tool to achieve meaningful analysis by simulations. As a first result, we could study the sensitivity of the AO system to the different mis-registrations, showing that the most critical parameters are the shifts and rotation. 

%
%
This work will be very relevant for the future ELT as all the SCAO modules of the first light instruments will most likely include a pyramid WFS in their design. The next step will be to consider the constraints of this telescope for the AO calibration. For instance, with no calibration source, a whole reference interaction matrix may not be available as it would most probably have to be acquired on-sky. In that case, acquiring only a few WFS signals (selected to maximize the sensitivity to each type of mis-registrations) might be sufficient to provide an accurate estimation of the mis-registrations parameters. There is a need to study the number and type of signals required to adjust the model mis-registrations parameters. If these signals are acquired on-sky, the impact of the noise on the parameters estimation will also have to be investigated.

In this paper, we highlighted that the key ingredients to generate an accurate pseudo synthetic interaction matrix is the estimation of the mis-registrations parameters. There is a need to optimise the identification and especially the tracking of these parameters during the operation, if possible without injecting any perturbation. This had to be done taking into account the complexity of the Pyramid WFS (modal sensitivity and linearity dependent on the seeing conditions and on the WFS modulation). Many concepts have already been proposed to this purpose. They still have to be carefully evaluated. 

\section*{Acknowledgements}
This project has received funding from the European Union's Horizon 2020 research and innovation programme under grant agreement No 730890. This material reflects only the authors views and the Commission is not liable for any use that may be made of the information contained therein.




\bibliographystyle{mnras}
\bibliography{C:/Users/cheritier/Documents/Bibliography/biblio} 







\bsp	
\label{lastpage}
\end{document}